\documentclass[journal]{IEEEtran}

\usepackage{cite}
\usepackage[utf8]{inputenc}
\usepackage{amsmath,amssymb,amsfonts}
\usepackage{algorithmic}
\usepackage[ruled]{algorithm2e}
\usepackage{graphicx}
\usepackage{textcomp}
\usepackage{xcolor}
\usepackage{blindtext}
\usepackage{diagbox}
\usepackage{hhline}
\usepackage{booktabs}
\usepackage{multirow}
\usepackage{subfigure}
\usepackage{verbatim}
\usepackage{url}
\usepackage{makecell, rotating, adjustbox}
\usepackage{caption}

\usepackage{hyperref}
\hypersetup{
colorlinks=true,
linkcolor=blue,
anchorcolor=blue,
citecolor=blue}

\newcommand{\MYhref}[3][blue]{\href{#2}{\color{#1}{#3}}}%

\begin{document}
\title{Digital Twin for Networking: \\A Data-driven Performance Modeling Perspective}

\author{Linbo Hui, Mowei Wang, Liang Zhang, Lu Lu, and Yong Cui

\thanks{Linbo Hui, Mowei Wang, and Yong Cui are with Tsinghua University (Yong Cui is the corresponding author, e-mail \MYhref[black]{mailto:cuiyong@tsinghua.edu.cn}{cuiyong@tsinghua.edu.cn}); 
Liang Zhang is with Huawei Technologies Co., Ltd; Lu Lu is with China Mobile Communications Group Co., Ltd.}}

\maketitle

\begin{abstract}
Emerging technologies and applications make the network unprecedentedly complex and heterogeneous, leading physical network practices to be costly and risky.
The digital twin network (DTN) can ease these burdens by virtually enabling users to understand how performance changes accordingly with modifications.
For this ``What-if" performance evaluation, conventional simulation and analytical approaches are inefficient, inaccurate, and inflexible, and we argue that data-driven methods are most promising.
In this article, we identify three requirements (fidelity, efficiency, and flexibility) for performance evaluation.
Then we present a comparison of selected data-driven methods and investigate their potential trends in data, models, and applications.
Although extensive applications have been enabled, there are still significant conflicts between models' capacities to handle diversified inputs and limited data collected from the production network.
We further illustrate the opportunities for data collection, model construction, and application prospects.
This survey aims to provide a reference for performance evaluation while also facilitating future DTN research.
\end{abstract}


\section{Introduction}
%
%
%
%

\IEEEPARstart{T}{he} endless pursuit for high-throughput and low-latency urges emerging technologies (e.g., 5G, cloud computing, edge computing) to be employed. 
Meanwhile, high-performance network services, in turn, have spawned a batch of new applications (e.g., live streaming, virtual reality, cloud gaming).
These technologies and applications make the network unprecedentedly complex and heterogeneous, leading to costly and risky practices on the physical network.
In this context, a digital twin network (DTN)~\cite{zhou-nmrg-digitaltwin-network-concepts-07}
can significantly ease the practitioners' burdens.
DTN is a virtual representation of the physical communication network that continuously updates with the latter's performance, maintenance, and health status data.
Unlike typical digital twin technologies that replicate physical objects, DTN is primarily concerned with the abstraction of network states and behaviors.
DTN aims to build digital twins for universal communication networks and is not restricted to specific applications or contexts.

Network operators often desire to develop optimization techniques to improve network performance, which mainly involves configuration tunning and new-policy exploration.
New configurations and policies must be fully verified before being deployed in physical networks.
DTN can be used as a safe and cost-efficient environment for performance evaluation. 
Operators may explore and verify their new techniques in DTN, avoiding complicated and risky operations on physical networks.
The optimization and other scenarios (\S II.A) demand performance evaluation for ``What-if'' scenarios~\cite{tariq2008answering}, which means the DTN can tell what the network performance is if there are alterations in influencing factors (e.g., traffic volumes, device configurations, routing schemes, topologies).

Network performance evaluation has attracted researchers' interest for decades. 
Experiments (e.g., A/B testing) and measurements are two techniques for performance evaluation with physical production networks, which are both high-risk, high-overhead, and high-complexity for ``What-if'' scenarios.
In virtual environments, simulation and modeling are two fundamental approaches for ``What-if'' performance evaluation.
Network simulators (e.g., NS-2, NS-3, OMNet++) process virtual packets under pre-defined mechanisms (e.g., congestion control algorithms, queueing policies) and generate performance metrics (e.g., throughput, delay, loss rate), which allow the collection of arbitrary information without impacting system behavior.
Such packet-level simulators are delicately designed and tightly coupled, leading to \textbf{inefficient} execution.
The 3-4 orders of magnitude slower than real-time~\cite{zhang2021mimicnet} determine that current simulators are unacceptable for DTN.
Modeling is much different from simulation, which directly establishes the relationships between influencing factors and performance metrics.
Conventional analytical modeling methods (e.g., network calculus, queuing theory) adopt Poisson Process to simplify the packet arrival and departure process, which can not describe the incast~\cite{geng2019simon} and leading to \textbf{inaccurate} estimation.
Simulators are heavy, while analytics are formalistic, and both are \textbf{inflexible} towards rapid and continuous network evolution.

With the renaissance of machine learning in recent years, data-driven techniques, especially neural networks (NNs), seem promising for DTN's performance evaluation.
Data-driven methods usually pre-define a series of possible mapping functions and utilize abundant data to determine a set of parameters (i.e., training) that accurately map the influencing factors to performance metrics.
The mapping function is able to describe various relationships, which provides flexibility for modeling complex network mechanisms.
Trained data-driven models are lightweight and can efficiently generate outputs with only one-time forward computation.
Researchers have developed specialized structures for specific tasks (e.g., convolutional neural networks for computer vision, recurrent neural networks for natural language processing) and applied techniques (e.g., regularizations, dropout) to solve the overfitting and generalization problems, making trained models reliable.
With these advantages, data-driven methods can help to mitigate the inefficiency, inflexibility, and inaccuracy issues.

In this article, we identify three requirements (fidelity, efficiency, and flexibility) for network performance evaluation by investigating four typical network scenarios.
Then we make a comprehensive comparison of selected data-driven performance models on data, models, and applications.
The data utilized are diversified and mainly collected from simulation environments.
The models develop from the classical-method stage, vanilla-NN stage to the customized-NN stage and strive to meet three requirements.
Though enabled extensive applications, models' practical usages are still rare, which indicates the significant conflicts between models' powerful expression abilities and the shortage of practical data from the production environment.
Based on the above, we further describe opportunities and challenges for performance evaluation from data collection, model construction, and application prospects.
DTN is undergoing rapid development, and performance evaluation is essential for DTN's construction.
This article surveys network performance evaluation from a data-driven perspective.
We hope that this survey will not only serve as a favorable reference for performance evaluation but also facilitate future research towards DTN.

\section{Requirements for Performance Models}

\subsection{Scenarios and Requirements}

\begin{figure}[t]
    \centering
    \includegraphics[width=0.48\textwidth]{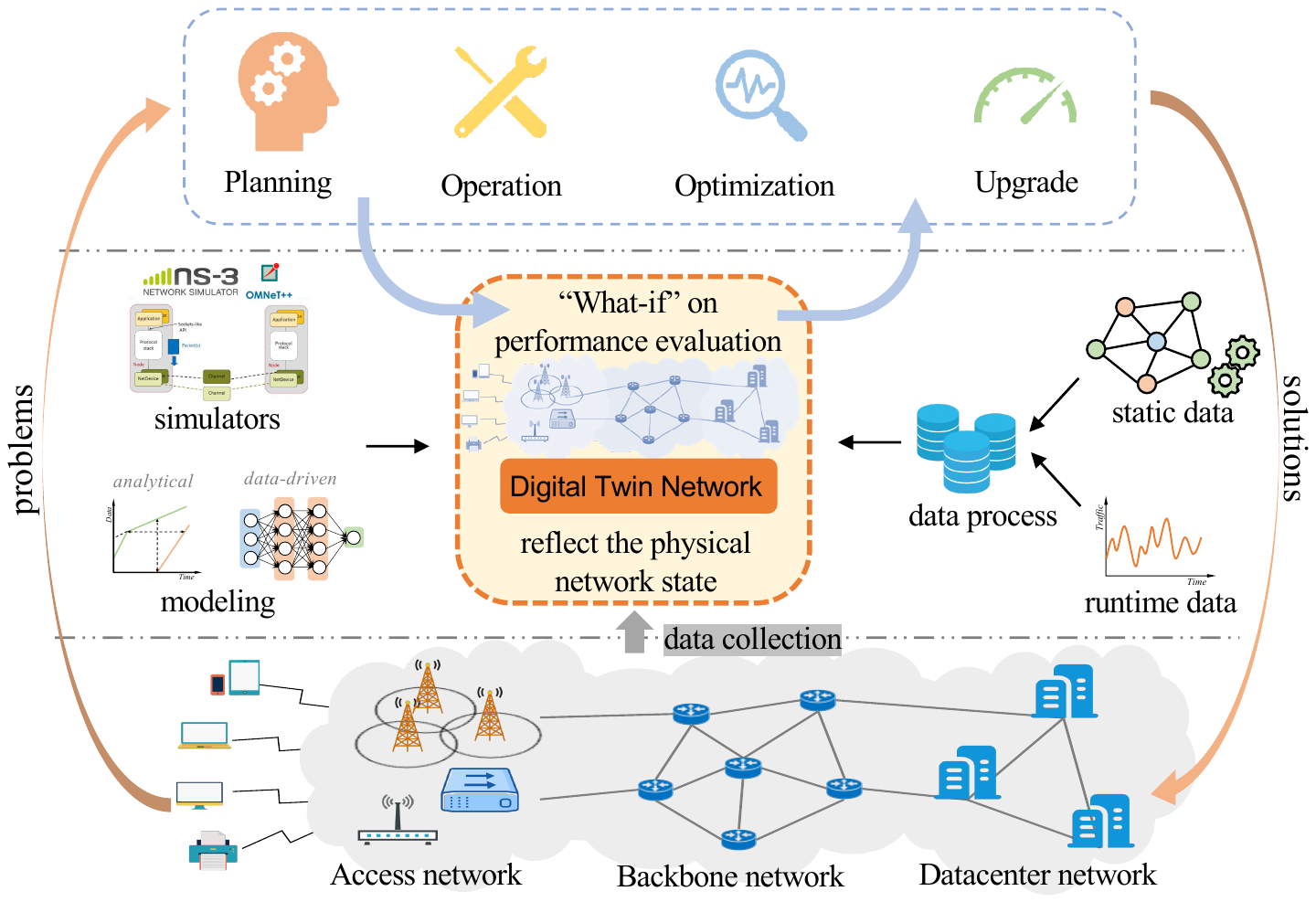}
    \caption{The Digital Twin Network Architecture.}
    \label{DTN-arch}
\end{figure}

Researchers have proposed a general DTN architecture~\cite{zhou-nmrg-digitaltwin-network-concepts-07} including three layers, as Figure~\ref{DTN-arch} shows.
It needs to periodically collect the static data (e.g., topology, configurations) and continuously collect the runtime data (e.g., link utilization, traffic volume) and populate them into DTN.
DTN then reconstructs the internal relations of collected data to represent the physical network state, which enables the ``What-if'' ability.
DTN will benefit network practices by providing a real-time and zero-risk performance evaluation environment.
We investigate four typical network scenarios of planning, operation, optimization, and upgrade to identify specific requirements for performance models.
\begin{itemize}
    \item In the planning scenario, designers need to ensure the planning network's overall performance meets the requirements of the given topology, configurations, and demand traffic.
    The model must generate performance results under different topologies, configurations, and traffic loads, which requires the model to be accurate on various inputs combinations. 
    
    \item When in operation, engineers hope to know about the real-time performance changes and quickly respond to potential anomalies.
    These anomalies must be quickly located or detected once they appear.
    Real-time monitoring and anomaly detection require the model to efficiently depict the physical network performance.
    
    \item Network optimization usually involves configurations tuning and new-policy exploration.
    Configurations and policies must be fully verified before being deployed in practice.
    The model is an ideal zero-risk environment to explore schemes and evaluate their performances under various scenarios before deployment.
    
    \item Network upgrade often includes topology changes and link expansions. 
    Operators may wonder how to change the topology and where to expand the bandwidth under a limited resource budget.
    The model needs to evaluate the performance under changed topologies and link capacities and tell where the bottlenecks are to maximize the upgrade effectiveness.
    
\end{itemize}

From the above respects, there are three requirements (i.e., fidelity, efficiency, and flexibility) that a performance model must strive to achieve.
Fidelity is the basic requirement that ensures accuracy for all scenarios. Efficiency is essential for operation and optimization because of real-time and frequent performance evaluation.
Flexibility mainly means the model can evaluate performance with network changes, facilitating network planning, optimization, and upgrade.
We further elaborate on three requirements as follows.

\subsection{Requirements Elaboration}
\subsubsection{Fidelity}
The fidelity shows how accurate the metrics from performance models are with the physical network.
We divide the fidelity into three levels (i.e., long-term, short-term, one-to-one) from the temporal perspective.
Long-term represents statistic results over a period of time, while short-term describes the detailed changes in time slots, and one-to-one precisely depicts every packet of the physical network.
Long-term \textbf{steady} evaluation requires the model to reasonably abstract the complex mechanisms, such as RouteNet~\cite{rusek2020routenet} can represent arbitrary routing schemes.
It further requires describing network conditions and \textbf{temporal} dependencies to represent the short-term process, such as xNet~\cite{Wang2205:XNet} learning the state transition function between time steps.
One-to-one means we must accurately model diverse mechanisms' influence on each packet.
MimicNet~\cite{zhang2021mimicnet} keeps end-hosts function and models network clusters' effects on packets, which takes the first step towards one-to-one modeling.

\vspace{0.1in}

\subsubsection{Efficiency}
Efficiency stands for two parts.
One is that performance models can be faster than the real-time physical network.
The other is that models should be easy to deploy and consume rational resources.
Faster than the physical enables users to forecast performance and react in advance to tackle potential anomalies.
Unlike simulators, it needs to simplify complex mechanisms and discard unnecessary details (e.g., packet payload, switching process) to speed up the evaluation.
There is often a trade-off between the modeling granularity (e.g., packet-level~\cite{zhang2021mimicnet}, flow-level~\cite{li2020traffic, Wang2205:XNet}, path-level~\cite{xiao2018deep, rusek2020routenet}) and speed, which are determined by the target problem settings.
The model needs to consider the trade-off and serve as a cost-effective backend to evaluate performance metrics.

\vspace{0.1in}

\subsubsection{Flexibility}
The network is evolving rapidly.
When topology, configurations, or mechanisms change, the model must accurately generate performance metrics as well.
We hope that the model can be iteratively upgraded towards new mechanisms, which means the model must be flexible.
There usually remain unchanged parts in a changing system, inspiring us to leverage layered or modular philosophies with the flexibility problem.
Layers and modules are decoupled but orchestrated to function in one model, where we can separately construct every module. 
New mechanisms will not affect the whole model but only related modules, which can be either upgraded or replaced.
Flexibility will accelerate the process of building an accurate performance model.

\begin{table*}
    \centering
    \captionsetup{}
    \caption{Comparison of Selected Data-driven Performance Models on Data, Models and Applications}
    \label{advance-table}
    \resizebox{\columnwidth*2}{!}{%
    \begin{tabular}{ccllll} 
    \toprule
    \multicolumn{1}{c}{\textbf{Groups}} & \begin{tabular}[c]{@{}c@{}}\textbf{Authors}\\\textbf{Year@Pubs}\end{tabular} & \multicolumn{1}{c}{\begin{tabular}[c]{@{}c@{}}\textbf{Problem }\\\textbf{Scopes}\end{tabular}} & \multicolumn{1}{c}{\begin{tabular}[c]{@{}c@{}}\textbf{Data }\\\textbf{(Source, Inputs and Outputs)}\end{tabular}} & \multicolumn{1}{c}{\begin{tabular}[c]{@{}c@{}}\textbf{Models }\\\textbf{(name, techniques, etc.)}\end{tabular}} & \multicolumn{1}{c}{\textbf{Applications}} \\ 
    \midrule
    \multirow{2}{*}[3ex]{\rotatebox[origin=c]{90}{Application Layer}} & \begin{tabular}[c]{@{}c@{}}Tariq, et al.\\2008@\\SIGCOMM~\cite{tariq2008answering}\end{tabular} & \begin{tabular}[c]{@{}l@{}}estimate service \\ response time\end{tabular} & \begin{tabular}[c]{@{}l@{}}\textbf{S}: Google's global web-search CDN\\\textbf{I}: tcpdump data with many features (e.g., \\timestamp, region, RTT, response time)\\\textbf{O}: service response time distribution\end{tabular} & \begin{tabular}[c]{@{}l@{}}WISE uses Causal Baysian Network\\to learn the causal structure and \\applies statistical intervention to \\predict the response time\end{tabular} & \begin{tabular}[c]{@{}l@{}}evaluate the response\\time under “What-If" \\scenarios\end{tabular} \\ 
    \cline{2-6}
     & \begin{tabular}[c]{@{}c@{}}Sun, et al.\\2016@\\SIGCOMM~\cite{sun2016cs2p}\end{tabular} & \begin{tabular}[c]{@{}l@{}}predict\\throughput\end{tabular} & \begin{tabular}[c]{@{}l@{}}\textbf{S:}iQIYI's operational CDN platform\textbf{}\\\textbf{I: }clientIP, ISP, AS, city and server\\\textbf{O: }throughput\end{tabular} & \begin{tabular}[c]{@{}l@{}}CS2P learns the parameters of Hidden \\Markov Model (HMM) via expectation-\\maximization (EM) algorithm\end{tabular} & \begin{tabular}[c]{@{}l@{}}predict throughput\\for video bitrate\\adaptation\end{tabular} \\ 
    \cline{1-6}
    \multirow{6}{*}[-17ex]{\rotatebox[origin=c]{90}{Tranport Layer}} & \begin{tabular}[c]{@{}c@{}}Mirza, et al.\\2010@ToN~\cite{mirza2010machine}\end{tabular} & \begin{tabular}[c]{@{}l@{}}predict TCP\\throughput\end{tabular} & \begin{tabular}[c]{@{}l@{}}\textbf{S}: laboratory WAN testbed\\\textbf{I}: transfer size and path properties (e.g., \\queuing delays, loss, available bandwidth)\\\textbf{O}: TCP throughput\end{tabular} & \begin{tabular}[c]{@{}l@{}}Support Vector Regression (SVR)\\with Radial Basis as kernel function,\\{loss funciton is $\epsilon$-insensitive loss} with\\L2 regularization\end{tabular} & \begin{tabular}[c]{@{}l@{}}predict end-to-end \\TCP throughput of\\wide area paths\end{tabular} \\ 
    \cline{2-6}
     & \begin{tabular}[c]{@{}c@{}}Nunes, et al.\\2014@JWCN~\cite{nunes2014machine}\\\end{tabular} & \begin{tabular}[c]{@{}l@{}}estimate RTT of \\TCP connection\end{tabular} & \begin{tabular}[c]{@{}l@{}}\textbf{S}: QualNet simulation\\\textbf{I}: three hyperparameters, measured RTT\\\textbf{O}: next RTT\end{tabular} & \begin{tabular}[c]{@{}l@{}}online fixed-share experts learning\\with weights updated every trail,\\loss is a piecewise function\end{tabular} & \begin{tabular}[c]{@{}l@{}}implemented inLinux \\ kernel to estimate \\TCP RTT\end{tabular} \\ 
    \cline{2-6}
     & \begin{tabular}[c]{@{}c@{}}Geyer, 2019@\\Performance\\Evaluation~\cite{geyer2019deepcomnet}\end{tabular} & \begin{tabular}[c]{@{}l@{}}evaluate the\\performance of\\a topology\end{tabular} & \begin{tabular}[c]{@{}l@{}}\textbf{S}: from simulation\\\textbf{I}: graph with nodes of flows and queues\\\textbf{O}: throughput of TCP flows, end-to-end\\latencies of UDP flows\end{tabular} & \begin{tabular}[c]{@{}l@{}}DeepComNet uses Graph Neural\\Network (GNN) with Gated Recurrent \\Units, loss function is MSE\end{tabular} & \begin{tabular}[c]{@{}l@{}}predict average TCP\\flows bandwidths and\\UDP flows end-to-\\end latencies\end{tabular} \\ 
    \cline{2-6}
     & \begin{tabular}[c]{@{}c@{}}Suzuki, et al.\\2020@ICOIN~\cite{suzuki2020estimating}\end{tabular} & \begin{tabular}[c]{@{}l@{}}infer end-to-\\end delay\end{tabular} & \begin{tabular}[c]{@{}l@{}}\textbf{S}: Fluid-based Simulator\\\textbf{I}: some delays of node-pairs and node \\features (e.g., indicator and degree)\\\textbf{O}: delays of other node-pairs\end{tabular} & \begin{tabular}[c]{@{}l@{}}semi-supervised GCN learning with\\rectified linear unit andlogarithmic \\softmax classifier, loss function is\\negative log likelihood\end{tabular} & \begin{tabular}[c]{@{}l@{}}infer delays of other \\node pairs from \\measured delays at \\some nodes.\end{tabular} \\ 
    \cline{2-6}
     & \begin{tabular}[c]{@{}c@{}}Zhang, et al.\\2021@\\SIGCOMM~\cite{zhang2021mimicnet}\end{tabular} & \begin{tabular}[c]{@{}l@{}}model DCN\\clusters' effects\\on packets.\end{tabular} & \begin{tabular}[c]{@{}l@{}}\textbf{S}: OMNet++ simulation\\\textbf{I}: scalable features (e.g., number of racks\\per cluster, packet size, priority bits)\\\textbf{O}: packet latency in the cluster\end{tabular} & \begin{tabular}[c]{@{}l@{}}MimicNet's internal model uses \\LSTM model, loss fucntion is \\ Weighted-BCE with Huber, two \\losses are weighted\end{tabular} & \begin{tabular}[c]{@{}l@{}}scalable, faster, and\\tunable performance\\estimation for DCN\end{tabular} \\ 
    \cline{2-6}
     & \begin{tabular}[c]{@{}c@{}}Wang, et al.\\2022@\\INFOCOM~\cite{Wang2205:XNet}\end{tabular} & \begin{tabular}[c]{@{}l@{}}model network\\performance\end{tabular} & \begin{tabular}[c]{@{}l@{}}\textbf{\textbf{S}}: NS-3 simulation\\\textbf{\textbf{I}}: traffic, buffer size, ECN, topology, queue\\policy, routing scheme, etc.\\\textbf{\textbf{O}}: path-/flow-level delay/throughput, FCT\end{tabular} & \begin{tabular}[c]{@{}l@{}}xNet uses three NGN blocks to build\\the state transition model, L2 loss \\function is used\end{tabular} & \begin{tabular}[c]{@{}l@{}}online QoS monitoring,\\“What-if" simulation,\\offline planning\end{tabular} \\ 
    \cline{1-6}
    \multirow{5}{*}[-13ex]{\rotatebox[origin=c]{90}{Network Layer}} 
    & \begin{tabular}[c]{@{}c@{}}Xiao, et al,\\2018@NetAI~\cite{xiao2018deep}\end{tabular} & \begin{tabular}[c]{@{}l@{}}infer path delay\\and loss distri-\\bution\end{tabular} & \begin{tabular}[c]{@{}l@{}}\textbf{S}: testbeds of DCN and WAN\\\textbf{I}: ports' load matrix in every four seconds\\\textbf{O}: path delay and loss distribution in\\every minute\end{tabular} & \begin{tabular}[c]{@{}l@{}}Deep-Q uses a variational auto-\\encoder (VAE) enhanced by the \\long short term memory (LSTM),\\loss function is Cinfer-loss\end{tabular} & \begin{tabular}[c]{@{}l@{}}infer QoS metrics of\\given traffic matrix\end{tabular} \\ 
    \cline{2-6}
     & \begin{tabular}[c]{@{}c@{}}Mestres, et al.\\2018@\\BigDAMA~\cite{mestres2018understanding}\end{tabular} & \begin{tabular}[c]{@{}l@{}}model end-to-\\end delay\end{tabular} & \begin{tabular}[c]{@{}l@{}}\textbf{S}: OMNet++ simulation\\\textbf{I}: traffic matrix of various scenarios\\\textbf{O}: end-to-end delay matrix\end{tabular} & \begin{tabular}[c]{@{}l@{}}neural networks model with different\\parameters, loss function is MSE \\with L2 regularization\end{tabular} & \begin{tabular}[c]{@{}l@{}}model end-to-end \\delay as function of \\traffic matrix\end{tabular} \\ 
    \cline{2-6}
    & \begin{tabular}[c]{@{}c@{}}Wang, et al.\\2018@\\SIGMETRICS~\cite{wang2018neural}\end{tabular} & \begin{tabular}[c]{@{}l@{}}evaluate the \\performance of\\DCN topology\end{tabular} & \begin{tabular}[c]{@{}l@{}}\textbf{\textbf{\textbf{\textbf{S}}}}:customized flow-level simulator\\\textbf{\textbf{\textbf{\textbf{I}}}}: demand traffic matrix and topology\\configuration\\\textbf{\textbf{\textbf{\textbf{O}}}}: performance metrics score\end{tabular} & \begin{tabular}[c]{@{}l@{}}xWeaver's scoring module uses two\\convolutional neural networkwith a\\fully-connected neural network, loss\\function is not mentioned\end{tabular} & \begin{tabular}[c]{@{}l@{}}evaluate and infer\\the propertopology\\of given traffic\end{tabular} \\ 
    \cline{2-6}
     & \begin{tabular}[c]{@{}c@{}}Rusek, et al.\\2020@JSAC~\cite{rusek2020routenet}\end{tabular} & \begin{tabular}[c]{@{}l@{}}model the path\\delay, jitter or \\loss\end{tabular} & \begin{tabular}[c]{@{}l@{}}\textbf{S}: OMNet++ simulation\\\textbf{I}: topology, routing schemes, link capacities,\\and path steady traffic\\\textbf{O}: path mean delay, jitter and loss\end{tabular} & \begin{tabular}[c]{@{}l@{}}RouteNet uses Message Passing\\Neural Network (MPNN) framework,\\loss is negative log likelihood\end{tabular} & \begin{tabular}[c]{@{}l@{}}QoS-aware routing \\optimization, budget-\\constrained network \\upgrade\end{tabular} \\ 
    \cline{2-6}
     & \begin{tabular}[c]{@{}c@{}}Li, et al.\\2020@CN~\cite{li2020traffic}\end{tabular} & \begin{tabular}[c]{@{}l@{}}model FCT \\in DCN\end{tabular} & \begin{tabular}[c]{@{}l@{}}\textbf{S}: DiffservNetwork simulator\\\textbf{I}: flow size and start time, ToS, protocol, \\bandwidth, etc.\\\textbf{O}: FCT\end{tabular} & \begin{tabular}[c]{@{}l@{}}GNN model (three Graph Networks\\blocks as encoder, core, decoder),\\loss function is MSE\end{tabular} & \begin{tabular}[c]{@{}l@{}}flow routing and\\scheduling, topology\\management for\\trafficoptimization\end{tabular} \\
    \bottomrule
    \end{tabular}}
\end{table*}

\section{Selected Performance Models}

\subsection{Advances Overview}
Network researchers have shown great interest in data-driven performance modeling in the past ten years.
We select some representatives and make a comprehensive comparison of problem scopes, data, models, applications in Table~\ref{advance-table}.
They are divided into three groups by applied layers.
Existing approaches are developed under different performance evaluation tasks where metrics include delay, throughput, loss rate, flow completion time (FCT), etc.
Some works focus on performance inference or estimation, and others focus on performance prediction.
There are no significant differences between the two focuses, as they just mean the evaluation of current or future performance.

Data, model, and application are three main aspects of a data-driven approach.
Data is essential to train the model, which is distinct from conventional analytical or simulation approaches. 
To some extent, the available data in quantity and quality limits how accurate a data-driven model will perform.
Meanwhile, models should be reasonably designed, and a well-designed model can efficiently extract variables' relations to construct itself.
Researchers have also developed specific techniques on model architecture and loss function to obtain more accurate results.
Most models are not only applied for targeted tasks but also leveraged for other scenarios, demonstrating the broad prospects of performance models.

\begin{figure}[t]
    \centering
    \includegraphics[width=0.45\textwidth]{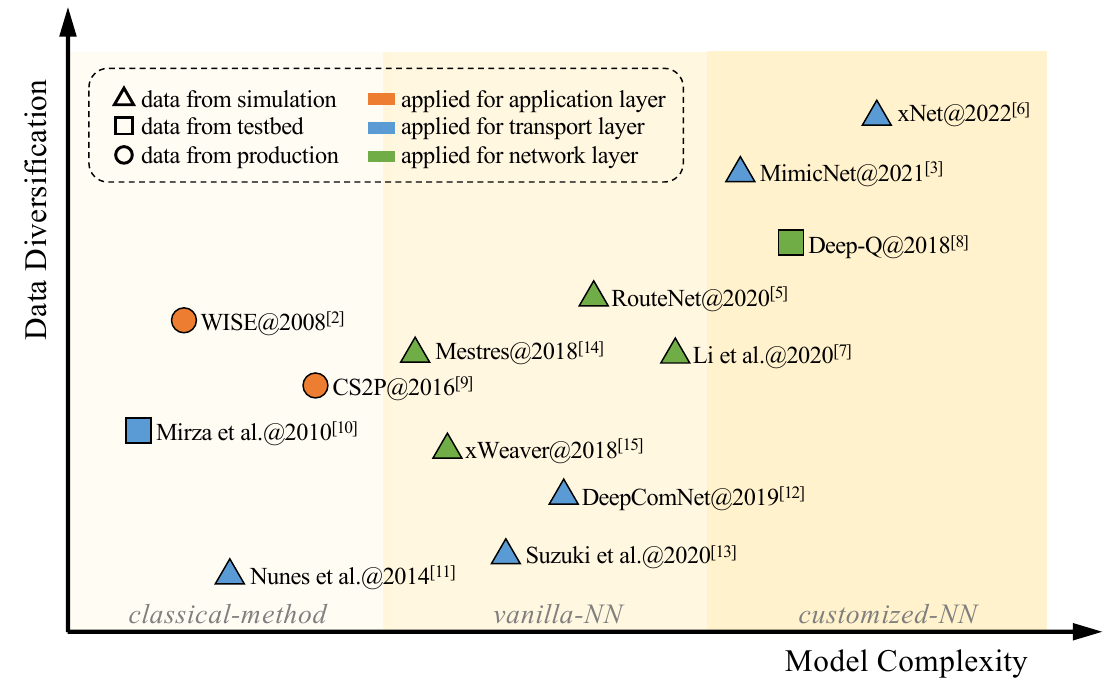}
    \caption{Qualitative comparison on data diversification, model complexity and application scenarios of selected data-driven performance models}
    \label{data-model}
\end{figure}

From an overview, we summarise potential trends in data, models, and applications.
The data utilized are diversified but mainly acquired from the simulation environment, indicating the major conflict between models' ability to handle complex inputs and the shortage of data collected from the production network.
Diversified data can yield better fidelity and practical applications.
Further, the leveraged techniques are advancing with machine learning development, resulting in more delicate models.
With the stronger abstraction and powerful expressiveness, models' fidelity and flexibility are improved.
At last, these models have enabled extensive application scenarios for different layers, mainly including network, transport, and application layers.
The lower the layer, the more detailed modeling abstractions are required. 
With the advancement of modeling techniques, there is roughly a tendency from higher to lower layer scenarios.
Researchers also realized the great potential of performance models and utilized them to solve problems that seemed to be challenging in the past, such as modeling the routing schemes~\cite{rusek2020routenet}.
Here we give a qualitative comparison of data diversification, model complexity, and application scenarios of selected models as Figure~\ref{data-model} shows.

\subsection{Data Source, Inputs and Outputs}

\subsubsection{Data source}
Data source mainly involves simulation, testbeds, and production, where the measurement difficulty sharply increases.
We notice that most of the selected works \cite{nunes2014machine, wang2018neural, mestres2018understanding, rusek2020routenet, suzuki2020estimating, li2020traffic, zhang2021mimicnet, Wang2205:XNet,geyer2019deepcomnet} are collecting data from the simulation environment.
They often leverage packet-level or flow-level simulators, which are costless and relatively easy to deploy.
Though various data are convenient to collect in such a restrained environment, the data fidelity may deviate from production.
Also, large-scale simulations with packet-level simulators may need a long time to complete.

Some works~\cite{mirza2010machine, xiao2018deep} build testbeds to obtain data.
Despite the high cost, testbeds are closer to the real environment and have fast speed.
Researchers can collect abundant clean data from controlled testbeds.
However, both simulators and testbeds have problems with the traffic model.
The flow patterns and distributions are hard to describe because of complex and diverse traffic behaviors.
Without reliable traffic models, the credibility of collected data will reduce.

Directly measuring data from production networks is wonderful, where simulation time and traffic model are not problems anymore.
WISE~\cite{tariq2008answering} used tcpdump data from Google's global web-search content delivery network (CDN), and Sun et al.~\cite{sun2016cs2p} used a proprietary dataset of HTTP throughput measurement from the operational platform of iQIYI.
Real data can bring higher credibility and enables the performance models to be deployed for practical usage.
However, collecting plenty of consistent and clean data from the production environment is still costly.
Specialized equipment and techniques are often needed to take measurements.
Even so, some metrics (e.g., queue length~\cite{geng2019simon}) can not be directly measured.
High-precision time synchronization on devices also hinders accurate measurements of time-related metrics.

\subsubsection{Inputs} 

Inputs are various for different tasks.
Some works~\cite{tariq2008answering, mirza2010machine, sun2016cs2p, rusek2020routenet, li2020traffic} take multiple discrete or continuous variables as inputs.
WISE~\cite{tariq2008answering} inspected a tcpdump dataset and found that a series of variables (e.g., timestamp, number of sent packets, client region, round-trip time (RTT), response time) are responsible for service response time.
Mirza et al.~\cite{mirza2010machine} proposed that incorporating path properties (e.g., queuing delays, loss, available bandwidth) as inputs could improve history-based TCP throughput prediction.
CS2P~\cite{sun2016cs2p} picked a given set of features from all possible features (e.g., client IP, ISP, AS, city, server) combinations to aggregate session clusters.
RouteNet~\cite{rusek2020routenet} utilized path steady traffic under various topologies, routing schemes, and link capacities to infer the path delay and loss rate.
Li et al.~\cite{li2020traffic} took flow size, start time, type of service (ToS), protocol, and link bandwidth to model flow completion time (FCT).
Though it introduces a heavier burden, diversified inputs will yield more accurate outputs and stronger generalization ability.

There are also works~\cite{nunes2014machine, wang2018neural, xiao2018deep, mestres2018understanding,geyer2019deepcomnet, suzuki2020estimating} taking fewer features as inputs.
Such works often focus on given scenarios and tackle specific tasks.
Nunes et al.~\cite{nunes2014machine} and Suzuki et al.~\cite{suzuki2020estimating} only used measured metrics (RTT/delays to some nodes) to estimate homogeneous metrics (next RTT/delays to other nodes).
xWeaver\cite{wang2018neural}, Deep-Q~\cite{xiao2018deep}, and Mestres et al.~\cite{mestres2018understanding} preferred to use traffic matrixes to model end-to-end performance matrixes, where topology and other configurations were specified.
DeepComNet~\cite{geyer2019deepcomnet} employed lower-level features (graph representation of topology and flows) to evaluate the average performance of network topology.
Fewer input features can reduce the measurement costs and enable fast execution, improving efficiency under specific settings.

\subsubsection{Outputs} 

Outputs are performance metrics (e.g., RTT, delay, throughput, loss rate, FCT) with fewer dimensions than inputs.
The majority of the works~\cite{mirza2010machine, sun2016cs2p, wang2018neural, mestres2018understanding, geyer2019deepcomnet, rusek2020routenet, suzuki2020estimating, li2020traffic} aim to evaluate long-term average performance under steady-state.
Mirza et al.~\cite{mirza2010machine} and CS2P~\cite{sun2016cs2p} both predicted throughput in a period.
xWeaver~\cite{wang2018neural} and DeepComNet~\cite{geyer2019deepcomnet} both estimated the given topology's steady performance.
Mestres et al.~\cite{mestres2018understanding} and RouteNet~\cite{rusek2020routenet} both evaluated path mean delay, which reflected the overall path performance under steady traffic.
Suzuki et al.~\cite{suzuki2020estimating} inferred end-to-end delays of node pairs under persistent TCP flows.
Li et al.~\cite{li2020traffic} modeld FCT in DCN.
Above long-term performance evaluation facilitated decision-making for target tasks.

Unlike the average metrics above, generating metrics distributions enables users to evaluate the performance from a probabilistic point.
WISE~\cite{tariq2008answering} estimated service response time distribution under ``What-if'' scenarios and could provide the percentage guarantee for service level agreement (SLA) requirements.
Deep-Q~\cite{xiao2018deep} verified that some quality of service (QoS) metrics are not a scalar but a random variable.
It inferred path delay and loss distributions over time intervals.
Deep-Q also provided performance changes over time, which enabled dynamic performance evaluation from a temporal dimension.
Nunes et al.~\cite{nunes2014machine} estimated the next RTT with past measured RTT.
xNet~\cite{Wang2205:XNet} enabled temporal prediction on QoS inference and FCT.
MimicNet~\cite{zhang2021mimicnet} modeled latency on each packet.
The temporal metrics reflect performance changes over time, enabling more valuable applications (e.g., QoS monitoring, anomalies detection, exploring policies).

\subsection{Model Selection and Customization}
From an overview of model construction, we divide recent advances into three stages: classical-method stage, vanilla-NN stage, and customized-NN stage.

\subsubsection{classical-method stage}
In this stage, classical methods (e.g., Causal Bayesian Network (CBN), Support Vector Regression (SVR)) are adopted for performance modeling tasks.
Though with fixed and straightforward forms, such methods have shown good performance for specific tasks.
WISE~\cite{tariq2008answering} identified relevant features from vast variables and leveraged CBN to construct the causal structure of these features.
Then it applied statistical interventions to changed features to estimate response time under ``What-if'' scenarios.
Mirza et al.~\cite{mirza2010machine} leveraged SVR with Radial Basis kernel function to predict TCP throughput.
The loss function is $\epsilon$-insensitive loss with L2 regularization.
SVR has a solid theoretical foundation and is favored in practice for its excellent empirical performance.
Nunes et al.~\cite{nunes2014machine} used lightweight Experts Framework to perform online learning, which showed high speed with reasonable accuracy.
They developed a dedicated piecewise loss function to describe the error between prediction and measurement.
CS2P~\cite{sun2016cs2p} employed Hidden Markov Model (HMM) to capture the state-transition behaviors for similar clusters to predict throughout for video bitrate adaptation.

In this stage, classical data-driven techniques often have fixed forms and few empirical hyperparameters, where loss functions are easy to determine.
Based on this, there is not much space to design particular architectures for specific tasks.
Researchers need to abstract the problems and pay attention to feature engineering.
These models are very efficient and have high fidelity when properly applied, while they may not be flexible for network evolution.
With fixed forms, it is challenging for them to model complicated network mechanisms, leading to very limited applications.

\subsubsection{vanilla-NN stage}

With the development of computing technology and abundant available data, NNs bring data-driven techniques into a new era.
NNs are adopted for network performance modeling as well, and this enables the vanilla-NN stage.
xWeaver~\cite{wang2018neural} and DeepComNet~\cite{geyer2019deepcomnet} both leveraged NN techniques to evaluate the performance of a given topology.
xWeaver can explore optimal topology for given traffic, where a scoring module was used for evaluation.
The scoring module used two Convolutional Neural Networks (CNNs) to extract information from traffic and topology configuration separately, with a fully-connected NN to output the performance score.
DeepComNet represented flows and queues as graph nodes.
When a flow traverses a queue, an edge is connected with the flow node and the queue node.
It then leveraged Gated Graph Neural Network (GGNN) to evaluate topology.
Mestres et al.~\cite{mestres2018understanding} wondered whether NN could accurately model the delay as a function of the input traffic.
They evaluated NN with different hyperparameters (i.e., number of hidden layers, number of neurons per layer, the activation function, the learning rate, and the regularization parameter) and gave an affirmative answer.
Suzuki et al.~\cite{suzuki2020estimating} performed semi-supervised learning with Graph Convolutional Networks (GCN) to infer end-to-end delay.
They modeled the problem as a classification question and used a logarithmic softmax classifier.
RouteNet~\cite{rusek2020routenet} utilized Message Passing Neural Network (MPNN) to model the relationships between links and paths, outputting the path mean delay, jitter, and loss.
Li et al.~\cite{li2020traffic} abstracted flows as graph nodes and links as graph edges.
They input features of flows and links to the model, built with three Graph Network (GN) blocks, to output FCT.

The above works addressed problems with suitable NN techniques, where model structures and loss functions keep conventional.
Emerging NN methods are directly adopted for specific tasks.
Their inputs are usually structured (e.g., graph structure) and can not be directly processed by classical techniques.
These inputs have multiple dimensions, enabling better flexibility for long-term performance modeling.

\subsubsection{customized-NN stage}
Researchers have realized that detailed performance modeling (e.g., short-time and one-to-one) is more valuable and practical.
Deep-Q~\cite{xiao2018deep} first modeled path delay and loss distribution at time slots.
They combined a variational auto-encoder (VAE) with the long short-term memory (LSTM).
LSTM was used to extract information from sequences of traffic matrix, and VAE could generate metric distributions.
A specially-designed Cinfer-loss module could measure the error of predicted QoS distributions, which enabled efficient and accurate training.
MimicNet~\cite{zhang2021mimicnet} provided over two orders of magnitude speedup compared to regular simulation for DCN of thousands of servers.
It used a data-driven model to replace the complex and slow clusters in DCN while keeping the scalability, flexibility, and accuracy.
For accurately modeling effects (drop, latency, ECN, etc.) on packets, they developed a loss function with Weighted-Binary Cross-Entropy (BCE) and Huber loss.
xNet~\cite{Wang2205:XNet} provides a general approach to model the network characteristics of concern with graph representations and configurable GNN blocks.
It learns the state transition between time steps and rolls it out to obtain the entire fine-grained prediction trajectory.
xNet used three Networking Graph Networks (NGNs) to build the state transition model and applied loss for state transitions.

These works modeled the network at a dynamic view and provided higher fidelity, where inputs and outputs changed over time.
They designed delicate NN models and adopted domain knowledge for accurate modeling.

\subsection{Applications Scenarios}

In the beginning, there were not many applications for data-driven models.
Many works~\cite{tariq2008answering, mirza2010machine, nunes2014machine, sun2016cs2p, wang2018neural, mestres2018understanding} focused on single-metric modeling.
These works were designed for specific tasks and were only used in the targeted scenarios.
WISE~\cite{tariq2008answering} evaluated the service response time under the ``What-if'' scenarios.
Nunes et al.~\cite{nunes2014machine} used online learning to estimate the RTT of a TCP connection.
It could be implemented in the Linux kernel to improve the accuracy of RTT estimation.
Mirza et al.~\cite{mirza2010machine} and CS2P~\cite{sun2016cs2p} both predicted throughput under specific background for file transfer and video bitrate adaptation separately.
Mestres et al.~\cite{mestres2018understanding} verified that NN could accurately model end-to-end delay as a function of the traffic matrix.
They did not explore the use cases for applications.
xWeaver~\cite{wang2018neural} was dedicated to evaluating the performance of topology.
It could be used to infer the proper topology for given traffic.
Above all, we have seen very limited applications for these works.
There may be multiple reasons behind in aspects of data collection and modeling techniques.
Fewer inputs, single outputs, and specialized model is only valid for specific tasks.
Researchers did not consider applying these models for more scenarios.

With the advances of various NN techniques, models are becoming more powerful.
Extensive applications have been proposed for related works~\cite{xiao2018deep, geyer2019deepcomnet, rusek2020routenet, suzuki2020estimating, li2020traffic, zhang2021mimicnet, Wang2205:XNet}.
Deep-Q~\cite{xiao2018deep} hoped to infer performance metrics directly from traffic statistics in real-time.
It could also be used for performance optimization.
DeepComNet~\cite{geyer2019deepcomnet} was designed to evaluate the performance of topology.
Engineers could leverage DeepComNet to predict TCP bandwidth or UDP end-to-end latency for network planning.
RouteNet~\cite{rusek2020routenet} set an excellent example in applying the model for multiple use cases.
They modeled path mean delay, jitter, and loss, which are used for QoS-aware routing and budget-constrained network upgrades.
Li et al.~\cite{li2020traffic} also showed multiple applications for traffic optimization of flow routing and scheduling, topology management.
MimicNet~\cite{zhang2021mimicnet} enhanced the packet-level simulator and provided scalable, faster, and tunable performance evaluation for DCN.
xNet~\cite{Wang2205:XNet} proposed a modeling framework and demonstrated three use cases of online QoS monitoring, simulation of ``What-if'' scenarios, and network planning.
It is evident that data-driven models have more extensive applications than ever before.
Generalized data-driven methods have reduced the difficulties of traditional problems with accurate, flexible, and efficient modeling techniques.

\section{Challenges and Opportunities}

Data-driven performance evaluation is rapidly advancing, and multiple applications will benefit from it.
Despite the bright prospects, there are still many obstacles to conquer, with opportunities lying behind.
We conclude challenges and opportunities with respect to data collection, model construction, and application prospects as follows.

\subsection{Data Collection}

Available data for training strongly affects the accuracy and generalization of learning models.
Data from production networks often with higher value but the amount and types are limited.
Real data for training will introduce a practical model for deployment, such as WISE~\cite{tariq2008answering}.
At the same time, diverse data need to be collected under various configurations, which is unrealistic in the production network.
Simulation environments may be a good alternative to solve the problem of accuracy and diversification.
However, such simulations are often time-consuming.
In addition, there is often a gap between simulated data and real data, hindering trained models from applying to production.
Data-driven techniques have been widely developed in broad fields, where general datasets and benchmarks contribute a lot, while there are still no widely recognized datasets and benchmarks for networking.

Experiments, measurements and data-driven methods are all promising for these problems.
Engineers may need to utilize specialized techniques (e.g., high-precision time synchronization, in-network telemetry) and equipment (e.g., sketch-related) to accurately measure valuable data in production~\cite{geng2019simon}.
Meanwhile, various data-driven techniques are helping to mitigate the problem too.
Network domain knowledge can be imposed on a few measured data for augmentation.
Other learning techniques with fewer data are also proposed, such as few-shot learning and self-supervised learning.
Transfer learning may help to transform the simulation-data trained model into a practical model with little real data.
Data-driven networking is evolving rapidly, and we are confident that standard~\textbf{datasets and benchmarks} will be formulated and bring significant promotion to this exciting field.

\subsection{Model Construction}
Data-driven models need to reflect the complex mechanisms of the network.
There are both global and local, spatial and temporal relations of network entities, and they tangle together.
Congestion control algorithms, queueing policies, routing schemes, etc., all have essential impacts on network performance.
Congestion control algorithms function at flow-level but global, where both end-hosts and in-network information are utilized.
Queuing policies schedule packets in the local switching node, influencing the end-hosts behaviors.
Routing schemes manage flows at the path level, which has significant effects on traffic distribution.
In addition, these mechanisms are time-sensitive (i.e., current state changes will impact future states).
How to accurately model the global and local, spatial and temporal mechanisms are still challenging.
We also notice that data-driven models often consume many resources on data collection, training, tuning, etc.
It is not trivial to ultimately construct such a model from the blank.

We advocate using the modular or layered conception, and analytical-NN combined cogitation to ease the complex modeling problem.
The tangled mechanisms can be decoupled using modular or layered parts that may be built and updated independently.
Analytical methods can also be introduced to enable higher accuracy and efficiency, which are often derived from domain knowledge.
Current data-driven models are developed separately, while a \textbf{foundation model} might be quite beneficial.
Pre-trained foundation models can serve as a backend and we only need to fine-tune it for specific tasks.
The foundation model for performance evaluation is similar to Transformer for natural language processing, which can provide universal and basic modeling ability.
The challenges for constructing foundation models are the complicated model constructions and the universal abstraction of network mechanisms. 
A potential method is to describe the network as packets sequences, where various mechanisms can be viewed as impactions on the interval time between packets. 
We are yearning for all kinds of foundation models to appear in the near future.

\subsection{Applications Prospects}
Data-driven methods have advantages over conventional methods in efficiency, fidelity, and flexibility, which conducts great potential for the optimization of network configuration.
Search-based optimization policies will be promoted with faster evaluation speed.
What's more, temporal NN models can efficiently make sequential evaluations, thus enabling model-based control optimizations.
With the model faithfully evaluating networks' performance, the DTN can provide a high-fidelity dynamic environment.
Reinforcement learning (RL) techniques are often employed for network decision-making.
The RL agent needs to perceive states from the environment and make specific actions to maximize rewards.
The virtual dynamic DTN environment is an ideal playground for RL algorithms to safely explore new policies.
There are also emerging technologies (e.g., Artificial Intelligence for IT Operations (AIOps), Self-driving Networks) to alleviate engineers from heavy manual operation works, where DTN can serve as an exploration and verification environment.

\section{Conclusion}

Though many challenges ahead, we will ultimately achieve the DTN technology with continuous research, and we are currently taking the first step.
The DTN's essential feature is the ``What-if'' ability, and the performance modeling plays a critical role.
Conventional simulation and analytical approaches are inefficient, inaccurate, and inflexible, and we argue that data-driven methods are the most promising to build the performance model.
Researchers have proposed a few methods of data-driven performance evaluation, while systematic summaries are still missing.
This article surveys selected data-driven performance models from data, models, and applications perspectives.
The data utilized are mainly collected from simulation environments, and the models develop from the classical-method stage, vanilla-NN stage to customized-NN stage.
Though enabled extensive applications, models' practical usages are still rare, which indicates the significant conflicts between models' powerful expression abilities and the shortage of practical data from the production environment.
We believe that standard datasets and benchmarks will be formulated, and foundation models will appear to promote this exciting field.
We anticipate that this survey will not only serve as a favorable reference for performance evaluation but also facilitate future research towards DTN.

\section*{Acknowledgment}

This work is supported by NSFC (NO.62132009 and NO.61872211) and Tsinghua University-China Mobile Communications Group Co., Ltd Joint Institute. 
\bibliographystyle{IEEEtran}
\bibliography{reference}

\begin{IEEEbiographynophoto}{Linbo Hui}
  received the B.E. degree from North China Electric Power University, Beijing, China, in 2016. 
  He is currently studying toward his M.E. degree in the Department of Computer Science and Technology, Tsinghua University, Beijing, China. 
  His major research interest is machine learning for network modeling.
\end{IEEEbiographynophoto}

\begin{IEEEbiographynophoto}{Mowei Wang}
  received the B.E. degree in communication engineering from Beijing University of Posts and Telecommunications, Beijing, China, in 2017. 
  He is currently working toward his Ph.D. degree in the Department of Computer Science and Technology, Tsinghua University, Beijing, China. 
  His research interests are in the areas of data center networks and machine learning.
\end{IEEEbiographynophoto}

\begin{IEEEbiographynophoto}{Liang Zhang}
  received the Ph.D. degree from Southeast University, Nanjing,China, in 2010. 
  He is currently Vice-Director of Huawei AI4NET Lab and Director of Huawei DataCom AI Department. 
  His research interests include intelligent fault analysis, network traffic analysis and network optimization.
\end{IEEEbiographynophoto}

\begin{IEEEbiographynophoto}{Lu Lu}
  is currently the deputy director of infrastructural network technology department in China Mobile Research Institute and the vice chairman of ITU SG13 WP1. 
  Her research interest covers mobile core network, future network architecture, computing and network convergence etc. 
\end{IEEEbiographynophoto}

\begin{IEEEbiographynophoto}{Yong Cui}
  received the B.E. and Ph.D. degrees both in CS Department from Tsinghua University. He is currently a professor. He served or serves at the editorial boards on IEEE TPDS, IEEE TCC, IEEE Network and IEEE Internet Computing. He co-chaired ACM Sigcomm'19 in Beijing and an IETF working group. He published over 100 papers with several Best Paper Awards and a dozen of Internet standard documents (RFC). His research interests include the Internet architecture and the data-driven network
\end{IEEEbiographynophoto}

\IEEEpeerreviewmaketitle

\ifCLASSOPTIONcaptionsoff
  \newpage
\fi

\end{document}